\documentclass[letterpaper,10pt]{article} 

\usepackage[mode=buildmissing]{standalone} 
\usepackage{opticameet3} 
\usepackage{acro}

\usepackage{hyperref} 
\usepackage{url}
\usepackage{graphicx}
\graphicspath{../Figure/}
\DeclareGraphicsExtensions{.pdf,.jpeg,.png,.jpg, .svg}
\usepackage{wrapfig}
\usepackage{xcolor}
\usepackage[font=footnotesize]{caption}
\usepackage{makecell}
\usepackage{psfrag}
\usepackage{tikz}
\usepackage{subcaption}
\usepackage{pgfplots}
\pgfplotsset{compat=1.18} 
\usepgfplotslibrary{groupplots}
\usetikzlibrary{calc}
\newcommand{\gettikzxy}[3]{%
  \tikz@scan@one@point\pgfutil@firstofone#1\relax
  \edef#2{\the\pgf@x}%
  \edef#3{\the\pgf@y}%
}

\usepackage{amsmath}
\usepackage{amsfonts}
\usepackage{bm}
\usepackage{mathrsfs}
\usepackage{mathtools}

\usepackage{array}

\newlength\myindent
\newlength\myindentt
\usepackage{url}

\DeclareAcronym{AIR}{short=AIR,long= achievable information rate,}

\DeclareAcronym{AWGN}{short=AWGN,long= additive white Gaussian noise,}

\DeclareAcronym{ASE}{short=ASE,long= amplified spontaneous emission,}

\DeclareAcronym{BW-CMA}{short=BW-CMA,long= block-wise CMA,}
\DeclareAcronym{BER}{short=BER,long= bit error rate,}
\DeclareAcronym{BW-DDLMS}{short=BW-DDLMS,long= DDLMS,}
\DeclareAcronym{CMA}{short=CMA,long= constant modulus algorithm,}
\DeclareAcronym{CD}{short=CD,long= chromatic dispersion,}
\DeclareAcronym{CW}{short=CW,long= continuous wave,}
\DeclareAcronym{CPE}{short = CPE,long = carrier phase estimation ,}
\DeclareAcronym{DD}{short=DD,long= decision-directed,}

\DeclareAcronym{DD-Kabsch}{short=DD-{K}absch,long= decision-directed {K}absch,}

\DeclareAcronym{DD-Czegledi}{short=DD-{C}zegledi,long=  decision-directed {C}zegledi,}

\DeclareAcronym{DDLMS}{short=DDLMS,long= decision-directed least mean squares,}

\DeclareAcronym{DP}{short=DP,long= dual-polarization,}
\DeclareAcronym{DGD}{short=DGD,long=  differential group delay,}
\DeclareAcronym{DP-PDL}{short=DP-PDL,long= dual-polarization PDL,}

\DeclareAcronym{DSP}{short=DSP,long= digital signal processing,}

\DeclareAcronym{DOF}{short=DOF,long= degrees of freedom,}
\DeclareAcronym{EDFA}{short=EDFA,long=  erbium-doped fiber-amplifier,}
\DeclareAcronym{EOFC}{short=EO comb,long=  electro-optic frequency comb,}

\DeclareAcronym{GD}{short=GD,long= gradient descent,}

\DeclareAcronym{GN}{short=GN,long= Gaussian Noise,}

\DeclareAcronym{GMI}{short=GMI,long= generalized mutual information,}

\DeclareAcronym{iid}{short = i.i.d.,long = identically and independently distributed,}
\DeclareAcronym{ISA}{short = ISA,long = inverse scattering algorithm,}
\DeclareAcronym{JCP}{short = JCP,long = joint-channel processing ,}

\DeclareAcronym{LMS}{short=LMS,long= least mean square,}

\DeclareAcronym{LS}{short=LS,long= least-square error,}

\DeclareAcronym{LS-DDLMS}{short= LS-DDLMS,long= LS-DDLMS,}

\DeclareAcronym{LS-SW-ULS}{short= LS-SW-ULS,long = LS-SW-ULS,}

\DeclareAcronym{LO}{short=LO,long =local oscillator,}
\DeclareAcronym{MCMA}{short= MCMA,long= modified {CMA},}

\DeclareAcronym{MI}{short=MI,long= mutual information,}

\DeclareAcronym{MIMO}{short=MIMO,long=  multiple-input multiple-output,}

\DeclareAcronym{ML}{short= ML,long= maximum likelihood}

\DeclareAcronym{MMA}{short=MMA,long= multi-modulus algorithm,}

\DeclareAcronym{MMSE}{short=MMSE,long= minimum mean square error,}

\DeclareAcronym{MSP}{short=MSP,long= master-slave processing,}

\DeclareAcronym{NLI}{short=NLI,long= nonlinear impairments,}

\DeclareAcronym{pdf}{short=pdf,long= probability density function,}

\DeclareAcronym{PDL}{short=PDL,long= polarization-dependent loss,}

\DeclareAcronym{PDM}{short= PDM,long= polarization-division multiplexed,}

\DeclareAcronym{PM}{short= PM,long= polarization multiplexed,}

\DeclareAcronym{PM-16-QAM}{short=PM-$16$-QAM,long= polarization-multiplexed $16$ quadrature amplitude modulation,}

\DeclareAcronym{PMD}{short=PMD,long= polarization-mode dispersion,}

\DeclareAcronym{PN}{short=PN,long= phase noise,}

\DeclareAcronym{PTC}{short=PTC,long= polarization-time code,}

\DeclareAcronym{QAM}{short=QAM,long= quadrature amplitude modulation,}

\DeclareAcronym{QPSK}{short=QPSK,long= quadrature phase-shift keying,}

\DeclareAcronym{RDE}{short=RDE,long= radially directed equalizer,}

\DeclareAcronym{RF}{short=RF,long=radio frequency,}

\DeclareAcronym{RHS}{short=RHS,long=right-hand side,}

\DeclareAcronym{RA}{short= RA, long = reference-assisted}
\DeclareAcronym{SDM}{short=SDM,long=Space-division multiplexing,}

\DeclareAcronym{SNR}{short=SNR,long=signal-to-noise ratio,}

\DeclareAcronym{SOP}{short=SOP,long= state of polarization,}

\DeclareAcronym{SVD}{short= SVD,long= singular value decomposition}

\DeclareAcronym{SER}{short=SER,long= symbol error rate,}

\DeclareAcronym{SW-Kabsch}{short= SW-Kabsch,long= sliding window Kabsch,}

\DeclareAcronym{SW-LS}{short= SW-LS,long= sliding window least squares,}

\DeclareAcronym{SW-ULS}{short= SW-ULS,long= sliding window unitary least square error,}

\DeclareAcronym{WSS}{short=WSS,long=  wavelength selective switches,}
\DeclareAcronym{WDM}{short=WDM,long=  wavelength-division multiplexing,}
\DeclareAcronym{WD}{short = WD,long = Wrapped Diagonal,}

\DeclareAcronym{WDT}{short = WDD,long = wrapped diagonal distribution,}
\DeclareAcronym{RAT}{short = RAD,long = reference-assisted distribution,}
\DeclareAcronym{EKS}{short = EKS,long = extended {K}alman smoother,}

	\usepackage{theoremref}
	\usepackage{ntheorem}

\definecolor{myMagneta}{rgb}{1, 0, 1}
\newcommand{\sigmaPhi}{\sigma^2}
\newcommand{\sigmaPsi}{\varrho^2}

\newcommand{\mr}[1]{\mathrm{#1}}

\DeclareMathAlphabet\mathbfcal{OMS}{cmsy}{b}{n}

\newcommand\norm[1]{\left\lVert#1\right\rVert}

\newcommand\oversetpi[1]{\mathstrut\mkern-1mu#1\mkern-18mu\raise1.25ex%
 \hbox{$\scriptscriptstyle 2\pi$}\mkern3mu}

\renewcommand{\H}{\mr{H}} 

\newcommand{\R}{\mr{R}}
\newcommand{\T}{\mr{T}}

\newcommand{\E}{\mathbb{E}}

\DeclareMathSymbol{\shortminus}{\mathbin}{AMSa}{"39}

\newcommand\authormark[1]{\textsuperscript{#1}}
\begin{document}

\title{Learning to Extract Distributed Polarization Sensing Data from Noisy Jones Matrices} 

\vspace{-0.6cm}


\author{
Mohammad Farsi\authormark{1},
    Christian H\"{a}ger\authormark{1}, Magnus Karlsson\authormark{2}, and
    Erik Agrell\authormark{1}
}
\address{\authormark{1}Dept.~of Electrical Engineering, Chalmers Univ.~of Technology, Sweden,\\
\authormark{2}Dept.~of Microtechnology and Nanoscience, Chalmers Univ.~of Technology, Sweden}

\email{\authormark{*}\textcolor{blue}{farsim@chalmers.se}} 

\vspace{-0.5cm}

\begin{abstract}
We consider the problem of recovering spatially resolved polarization information from receiver Jones matrices. 
We introduce a physics-based learning approach, improving noise resilience compared to previous inverse scattering methods, 
while highlighting challenges related to model overparameterization. 
\end{abstract}

\vspace{-0.03cm}

\section{Introduction}
The possibility of repurposing existing telecommunication fibers as environmental sensors has recently gained significant attention \cite{ip_using_2022}. 
Compared to established sensing approaches like distributed acoustic sensing, extracting polarization sensing data directly from adaptive equalizers in coherent communication receivers \cite{boitier_proactive_2017, Zhan2021, mecozzi_polarization_2021, mazur_field_2023} could have potential advantages, particularly in terms of cost, scalability, and compatibility with existing fiber infrastructure and equipment. 
An important open question is to what extent polarization sensing can provide distributed (i.e., spatially resolved) location information about external environmental events. 
Prior work makes relatively strong assumptions such as access to multiple optical paths \cite{castellanos_optical_2022}, per-span readouts \cite{simsarian_shake_2017}, or loop-back configurations \cite{charlton_field_2017}. 

In this paper, we study the problem of recovering spatially resolved polarization information directly from the estimated (frequency-dependent) Jones matrix at the receiver. 
A similar problem has been previously studied in \cite{harris_optical_1964, moller_filter_2000, noe_polarization-dependent_2015}. 
In \cite{harris_optical_1964, moller_filter_2000}, an \ac{ISA} is considered to recover the distributed \ac{DGD} profile from the overall system impulse response (i.e., the time-domain Jones matrix), essentially factorizing the system into a cascade of individual responses. 
The \ac{ISA} was extended in \cite{noe_polarization-dependent_2015} to account for \ac{PDL}, recovering both \ac{DGD} and \ac{PDL} profiles. 
In principle, the \ac{ISA} is exact for the settings considered in \cite{harris_optical_1964, moller_filter_2000, noe_polarization-dependent_2015}, assuming that the overall impulse response is known and noise-free. 

Our contributions are as follows. 
First, we show that the performance of the \ac{ISA} quickly deteriorates in the presence of noise. 
We then propose and investigate a physics-based learning approach \cite{Haeger2018ofc, Haeger2021jsac}, which directly parameterizes the underlying propagation model and jointly optimizes all parameters. 
We then show some examples where this approach can recover location information about time-varying polarization perturbations, even in the presence of noise. 
Finally, we discuss persisting limitations and challenges, in particular cases where model overparameterizations lead to ambiguities in interpreting the learned parameter configurations. 
We note that our learning approach is also related to work on distributed \ac{PMD} compensation \cite{Goroshko2016, Czegledi2017, Liga2018a, Bitachon2020b, Butler2021, jain_joint_2023, abu-romoh_equalization_2023, Liu2023ofc} and longitudinal path and \ac{PDL} monitoring \cite{sasai_digital_2022, eto_location-resolved_2022, may_receiver-based_2022, takahashi_dsp-based_2023}. 
However, \cite{Goroshko2016, Czegledi2017, Liga2018a, Bitachon2020b, Butler2021, jain_joint_2023, abu-romoh_equalization_2023, Liu2023ofc} do not investigate the feasibility of utilizing the optimized model representations for sensing or monitoring purposes and the approaches in \cite{sasai_digital_2022, eto_location-resolved_2022, may_receiver-based_2022, takahashi_dsp-based_2023} exploit additional nonlinear effects to enable spatial resolvability. 
The setting considered in this paper is different as we only assume access to the estimated (linear) Jones matrix. 
As such, the proposed approach is in principle fully compatible with existing polarization sensing approaches based on adaptive equalization. 

\section{System Model}
We study dual-polarization optical transmission accounting for \ac{PDL}, polarization rotations, and \ac{DGD}. 
The assumed channel model is based on \cite{noe_polarization-dependent_2015}.
In particular, the overall channel response at time $k$ and frequency $\omega_i$ is described by a $2\times 2$ complex-valued Jones matrix resulting from $N$ concatenated sections according to
\begin{align}
    \H(\omega_i;\theta_k) = \prod_{n=1}^{N} \Gamma(\gamma_{n}) \R\left(\phi_{n}(k),\psi_{n}(k)\right)\T(\omega_i), \label{eq:concat_rule1}
\end{align}
where we use the convention $\prod_{n=1}^{N} A_n = A_N A_{N-1}\cdots A_1$, $n\in\{1,\dots,N\}$ denotes the section index, 
 \begin{align}
    \!\!\Gamma(\gamma_n) = \begin{bmatrix}
        e^{\gamma_n/2} & 0\\
        0 & e^{-\gamma_n/2} 
    \end{bmatrix}, \,\,\R(\phi_n(k),\psi_n(k)) = \begin{bmatrix}
        \cos\phi_n(k) & \!\!\!\! je^{j\psi_n(k)}\sin\phi_n(k)\\
        je^{-j\psi_n(k)}\sin\phi_n(k) & \!\!\!\!\cos\phi_n(k) 
    \end{bmatrix}, \,\, \T(\omega_i) = \begin{bmatrix}
        1 & 0\\
        0 & e^{j\omega_i \tau} 
    \end{bmatrix}\label{eq:Gamma_matrix},
\end{align}
$\gamma_n$ is the \ac{PDL} extinction parameter, $\phi_n(k)$ and $\psi_n(k)$ are (possibly time-varying) rotation angles, $\tau$ is the \ac{DGD} per section, and 
$\theta_k =\left\{\gamma_n,\phi_n(k),\psi_n(k)\right\}_{n=1}^N$ denotes all channel parameters.
In addition to \cite{noe_polarization-dependent_2015}, we assume that an environmental perturbation introduces time variations in the rotation angles $\phi_n(k)$ and $\psi_n(k)$.
The time evolution is modeled as a random walk according to $\phi_n(k) = \phi_n(k-1) + \Phi_n(k)$ and $\psi_n(k) = \psi_n(k-1) + \Psi_n(k)$ where $\Phi_n(k)\sim\mathcal{N}(0,\sigmaPhi_n(k))$ and  $\Psi_n(k)\sim \mathcal{N}(0,\sigmaPsi_n(k))$. Here,
$\sigmaPhi_n(k)$ and $\sigmaPsi_n(k)$ are emulating the intensity of the environmental change, i.e., larger values correspond to more severe changes in the environment. 
Note that the model \eqref{eq:concat_rule1} assumes that the time scale of the dynamic polarization perturbation is much slower than the \ac{DGD} parameter $\tau$. 
 
In coherent receivers, the Jones matrix \eqref{eq:concat_rule1} is typically estimated through adaptive equalization. 
We therefore assume that we have access to a noisy version of the Jones matrix $\Tilde{\H}(\omega_i;\theta_k) = \H(\omega_i;\theta_k) + \mr{Z}_k(\omega_i)$, 
where $\mr{Z}_k(\omega_i)$ is a $2\times2$ matrix with zero-mean independent complex normal elements with variance $\sigma^2_{z}$ modeling the estimation error, i.e., $\E[Z_k(\omega_i)\circ Z_k(\omega_i)^\dagger] = \sigma^2_{z} \left( \begin{smallmatrix}
        1 & 1 \\
        1 & 1
    \end{smallmatrix}\right)$, where $\circ$ is the Hadamard product.    
The estimation noise variance $\sigma^2_{z}$ depends on various factors such as the choice of estimation algorithm and the channel \ac{SNR}, where better algorithms and higher \ac{SNR} lead to lower $\sigma^2_{z}$.

\section{Distributed Parameter Estimation Techniques}

In this section, we describe the \ac{ISA} and the proposed learning approach, both aiming at recovering the unknown parameters $\theta_k$ from $\tilde{\mr{H}}(\omega_i, \theta_k)$.\footnote{Open-source implementations are provided at \href{https://github.com/Mohammadfarsi1994/physics-based-distributed-polarization-sensing}{https://github.com/Mohammadfarsi1994/physics-based-distributed-polarization-sensing}.}

\noindent\emph{Inverse Scattering Algorithm:}
The \ac{ISA} proposed in \cite{noe_polarization-dependent_2015} operates under the assumption that the overall channel response is noiseless. 
The algorithm analytically calculates the values of the unknown parameters $\theta_k$ and
is performed iteratively in the time domain by first taking the inverse Fourier transform of $\tilde{\mr{H}}(\omega_i, \theta_k)$. 
The algorithm's procedure then calculates the unknown parameter values for section $N$, i.e., $\gamma_N$, $\phi_N(k)$, and $\psi_N(k)$, and subsequently employs these values to construct the equivalent total channel response for sections $1$ to $N-1$, effectively removing the last section from the model. By repeating these steps, all the subsequent parameters will be obtained.

\noindent\emph{Machine Learning Approach:}
We propose a physics-based machine learning approach, which directly parameterizes the propagation model \eqref{eq:concat_rule1} and jointly optimizes all its associated parameters. 
Denoting the set of trainable parameters as  $\hat{\theta}_k =\{\hat{\gamma}_n,\hat{\phi}_n(k),\hat{\psi}_n(k)\}_{n=1}^N$, we define the frequency-averaged cost function as
$\mathcal{L}(\hat{\theta}_k) = \sum_{i=1}^{L} \norm{\tilde{\H}(\omega_i;\theta_k)-\H(\omega_i; \hat{\theta}_k)}^2$, 
where $L$ is the number of frequency samples.
This loss function is then iteratively minimized by applying $M$ iterations of a gradient-descent optimizer with learning rate $\alpha$. 

\section{Numerical Results and Discussion}

For the numerical results, we consider $N=5$ sections, where, within each section, we sample the extinction parameter $\gamma_n$ from a uniform distribution within the range of $[0.07, 0.17]$ (equivalent to $[0.3, 0.7]$ dB \ac{PDL}). Furthermore, the initial rotation parameters $\phi_n(0)$ and $\psi_n(0)$ are sampled uniformly from $[-\pi, \pi]$. 
The \ac{DGD} parameter $\tau$ is normalized to $1$ and assumed to be known and constant across all sections. 
Note that this assumption is required for the \ac{ISA}, whereas the learning approach can potentially be extended to also learn unknown \ac{DGD} sections, similar to, e.g., \cite{Butler2021}. 
For the learning approach, the Adam optimizer is used with $M=300$ and $\alpha=0.05$, where all trainable parameters $\hat{\theta}_k$ are initialized to zero at $k=0$ and then tracked for $k>0$.
The frequency samples in the loss function correspond to $1/\tau$ sampling distance in the frequency domain with $L=N$, where we verified that increasing the number of frequency samples does not affect the results. 
Lastly, we consider a time-varying scenario where an environmental change within section $n=2$ at $k\in [15, 35]$ affects parameters $\phi_2(k)$ and $\psi_2(k)$ with $\sigmaPhi_2(k) = \sigmaPsi_2(k) = 0.1$ for $k\in[15, 35]$ and $\sigmaPhi_2(k) = \sigmaPsi_2(k) = 0$ elsewhere. 

\begin{figure}[t]
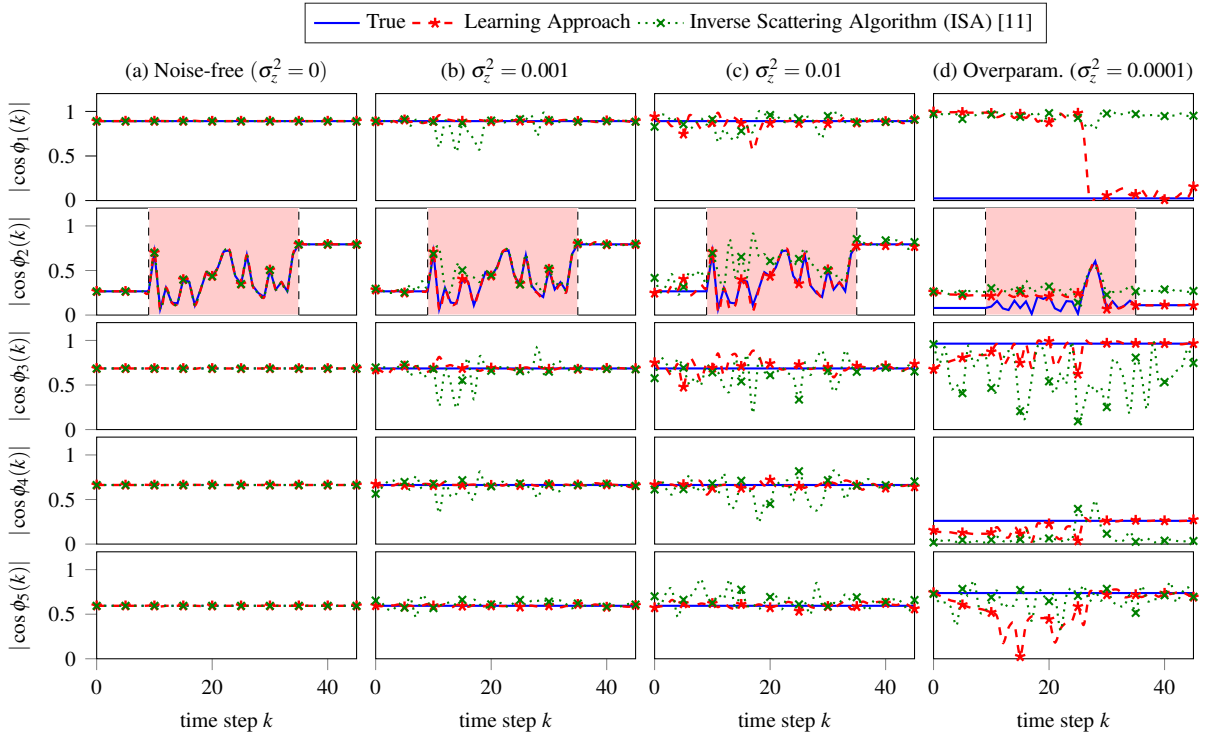

    \centering
\begin{tikzpicture}

\definecolor{darkgray176}{RGB}{176,176,176}
\definecolor{green}{RGB}{0,128,0}
\definecolor{lightgray204}{RGB}{204,204,204}
\newcommand\w{5}
\newcommand\h{3}
\newcommand\xmax{45}
\newcommand\xmin{0}
\newcommand\LineWidth{1.2}
\newcommand\trueLineWidth{0.8}
\newcommand\invLineWidth{0.8}
\newcommand{\plotTrue}{\addplot [semithick, blue, line width=\trueLineWidth pt]}
\newcommand{\plotNetwork}{\addplot [semithick, dashed, only marks, red,line width=0.9 pt,mark=star,mark repeat=5,smooth, mark options={fill=white, solid}]}
\newcommand{\plotInvscat}{\addplot [dotted, only marks, green,line width=0.8 pt, mark=x,mark repeat=5,smooth, mark options={solid}]}

\newcommand{\perturbation}{%
\addplot[black, semithick, <->] coordinates {(9,1.05) (35, 1.05)} node[midway, fill=white] {\scriptsize perturbation};
\draw[draw=white, fill=red!20, fill opacity=0.2] (9,0) rectangle (35, 1.5);
\addplot[black, dashed] coordinates {(9,0) (9, 1.5)};
\addplot[black, dashed] coordinates {(35,0) (35, 1.5)};
}%

\newcommand{\firstgplot}[2]{
\nextgroupplot[
width=\w cm,
height=\h cm,
anchor=north west,
at={($(#1 plots c1r1.north east) + (0.25cm,0)$)},
legend cell align={left},
legend style={at={(1,2)}, anchor=north west, legend cell align=left, align=left, font=\footnotesize, draw=white!15!black},
legend columns=3,
tick align=outside,
tick pos=left,
x grid style={darkgray176},
xmin=\xmin, xmax=\xmax,
xtick style={color=black,font=\footnotesize},
xlabel style ={color=black,font=\footnotesize},
ylabel style ={color=black,font=\footnotesize},
y grid style={darkgray176},
title = {#2},
title style ={color=black,font=\footnotesize, yshift=-0.2cm},
ymin=0, ymax=1.2,
ytick style={color=black,font=\footnotesize},
xtick style={color=white,font=\footnotesize},
xticklabels ={},
ytick style={color=white,font=\footnotesize},
yticklabels ={},
]
\pgfplotsset{every tick label/.append style={font=\footnotesize},}
}
\newcommand{\nextgplot}{
\nextgroupplot[
width=\w cm,
height=\h cm,
tick align=outside,
tick pos=left,
x grid style={darkgray176},
xmin=\xmin, xmax=\xmax,
y grid style={darkgray176},
xlabel style ={color=black,font=\footnotesize},
ylabel style ={color=black,font=\footnotesize},
ymin=0, ymax=1.2,
title style ={color=black,font=\footnotesize, yshift=-0.2cm},
xtick style={color=white},
xticklabels={},
ytick style={color=white},
yticklabels={},
]
\pgfplotsset{every tick label/.append style={font=\footnotesize},}
}
\newcommand{\lastgplot}{
\nextgroupplot[
width=\w cm,
height=\h cm,
tick align=outside,
tick pos=left,
x grid style={darkgray176},
xmin=\xmin, xmax=\xmax,
y grid style={darkgray176},
xlabel style ={color=black,font=\footnotesize},
ylabel style ={color=black,font=\footnotesize},
ymin=0, ymax=1.2,
title style ={color=black,font=\footnotesize, yshift=-0.2cm},
ytick style={color=white},
yticklabels={},
xlabel = {time step \(k\)},
]
\pgfplotsset{every tick label/.append style={font=\footnotesize},}
}

\input{Figs/random_phi_psi_change/noise_free}
\input{Figs/random_phi_psi_change/noise_0.001}
\input{Figs/random_phi_psi_change/noise_0.01}
\input{Figs/random_phi_change/failed_case_noise_0.0001}
\end{tikzpicture}
    \vspace{-0.4cm}
    \caption{Numerical results: (a)-(c) trackable channel realization with varying amounts of noise; (d) non-trackable realization.} 
    \vspace{-0.8cm}    \label{fig:numerical_results}
\end{figure}

Fig.~\ref{fig:numerical_results} compares the tracking capability of the \ac{ISA} and the learning approach by plotting $|\cos\phi_n(k)|$ for all sections in four different scenarios. The absolute value is used since two negative signs can potentially commute between different sections in \eqref{eq:concat_rule1}, effectively canceling each other and leading to sign ambiguity of the estimated angle parameters. 

\emph{Trackable cases:}
As depicted in Fig.~\ref{fig:numerical_results}(a), when the noiseless channel response is accessible, both methods are able to track the perturbation, pinpointing its location and timing. However, the situation changes when we examine Figs.~\ref{fig:numerical_results}(b) and (c), in which the channel response is subject to varying amounts of noise. In these cases, our proposed approach demonstrates improved noise resilience by successfully tracking and identifying the perturbation's location and timing, whereas the \ac{ISA} loses its capability to track or precisely determine the section affected by the environmental change.

\emph{Non-trackable cases:}
For certain realizations of the channel parameters $\theta_k$, 
there are situations where simultaneous changes in multiple parameters result in nearly identical channel responses $H(\omega_i; \theta_k)$ implying that it is possible to approximately describe $H(\omega_i; \theta_k)$ using fewer independent parameters. 
An example is shown in Fig.~\ref{fig:numerical_results}(d). Such cases manifest different behavior for the two algorithms. For the \ac{ISA}, numerical instabilities arise due to noise and neither the overall response nor the estimated parameters align with the actual model. For the learning method, the learned overall channel response $\mr{H}(\omega_i; \hat{\theta}_k)$ remains close to the actual response $\mr{H}(\omega_i; \theta_k)$ due to the employed loss function. However, since there are many parameter configurations that produce approximately the same response, the model is effectively overparameterized and no stable, unique solution can be found.
To address this issue, one possible approach would be to reduce the number of trainable parameters, for example by reducing the number of sections in the learned model, potentially sacrificing some localization accuracy. 
In this case, however, the ISA is no longer applicable and it becomes nontrivial to assess how well the learned distributed response matches the propagation model because it is not possible to compare on a parameter-by-parameter basis. 


\section{Conclusion}

In this paper, we considered the problem of estimating the distributed channel response from noisy Jones matrices. 
We showed that the previously proposed \ac{ISA} is relatively susceptible to noise. 
To address this limitation, we proposed a physics-based learning approach which exploits the propagation model and optimizes all parameters simultaneously, yielding promising results. 
We demonstrated successful location retrieval of a time-varying polarization perturbation in noisy conditions. 
Persisting challenges relate to cases of model overparameterization, making it ambiguous to relate the learned parameters to the actual perturbation in channel parameters. A potential avenue for future work involves modeling the actual channel response using a reduced number of sections, which may offer a solution to the overparameterization problem by trading off localization resolution.

\vspace{0.1cm}

\noindent\footnotesize{\textbf{Acknowledgements: }The work of C.~H\"ager was supported by the Swedish Research Council under grant no.~2020-04718.}

\vspace{-0.3cm}
\bibliography{references}

\begin{thebibliography}{10}
\newcommand{\enquote}[1]{``#1''}

\bibitem{ip_using_2022}
E.~Ip \emph{et~al.}, \enquote{Using {{Global Existing Fiber Networks}} for {{Environmental Sensing}},} {\protect\JournalTitle{Proc. IEEE}} \textbf{110}, 1853--1888 (2022).

\bibitem{boitier_proactive_2017}
F.~Boitier \emph{et~al.}, \enquote{Proactive {{Fiber Damage Detection}} in {{Real-time Coherent Receiver}},} in \emph{Proc. {{ECOC}},}  (2017).

\bibitem{Zhan2021}
Z.~Zhan \emph{et~al.}, \enquote{Optical polarization–based seismic and water wave sensing on transoceanic cables,} {\protect\JournalTitle{Science}} \textbf{371}, 931--936 (2021).

\bibitem{mecozzi_polarization_2021}
A.~Mecozzi \emph{et~al.}, \enquote{Polarization sensing using submarine optical cables,} {\protect\JournalTitle{Optica}} \textbf{8}, 788--795 (2021).

\bibitem{mazur_field_2023}
M.~Mazur \emph{et~al.}, \enquote{Field {{Trial}} of {{FPGA-Based Real-Time Sensing Transceiver}} over 524 km of {{Live Aerial Fiber}},} in \emph{Proc. OFC,}  (2023).

\bibitem{castellanos_optical_2022}
J.~C. Castellanos \emph{et~al.}, \enquote{Optical polarization-based sensing and localization of submarine earthquakes,} in \emph{Proc. OFC,}  (2022).

\bibitem{simsarian_shake_2017}
J.~E. Simsarian and P.~J. Winzer, \enquote{Shake {{Before Break}}: {{Per-Span Fiber Sensing}} with {{In-Line Polarization Monitoring}},} in \emph{Proc. OFC,}  (2017).

\bibitem{charlton_field_2017}
D.~Charlton \emph{et~al.}, \enquote{Field measurements of {{SOP}} transients in {{OPGW}}, with time and location [...],} {\protect\JournalTitle{Opt. Express}} \textbf{25}, 9689--9696 (2017).

\bibitem{harris_optical_1964}
S.~E. Harris \emph{et~al.}, \enquote{Optical {{Network Synthesis Using Birefringent Crystals}} [...],} {\protect\JournalTitle{J. Opt. Soc. Am.}} \textbf{54}, 1267--1279 (1964).

\bibitem{moller_filter_2000}
L.~Moller, \enquote{Filter synthesis for broad-band {{PMD}} compensation in {{WDM}} systems,} {\protect\JournalTitle{IEEE Photon. Technol. Lett.}} \textbf{12}, 1258--1260 (2000).

\bibitem{noe_polarization-dependent_2015}
R.~No\'e \emph{et~al.}, \enquote{Polarization-{{Dependent Loss}}: {{New Definition}} and {{Measurement Techniques}},} {\protect\JournalTitle{J. Lightw. Technol.}} \textbf{33}, 2127--2138 (2015).

\bibitem{Haeger2018ofc}
C.~Häger and H.~D. Pfister, \enquote{Nonlinear {{Interference Mitigation}} via {{Deep Neural Networks}},} in \emph{Proc. OFC,}  (2018).

\bibitem{Haeger2021jsac}
C.~Häger and H.~D. Pfister, \enquote{Physics-{{Based Deep Learning}} for {{Fiber-Optic [...]}},} {\protect\JournalTitle{IEEE J. Sel. Areas Commun.}} \textbf{39}, 280--294 (2021).

\bibitem{Goroshko2016}
K.~Goroshko \emph{et~al.}, \enquote{Overcoming performance limitations of digital back propagation due to polarization mode dispersion,} in \emph{Proc. ICTON,}  (2016).

\bibitem{Czegledi2017}
C.~B. Czegledi \emph{et~al.}, \enquote{Digital backpropagation accounting for polarization-mode dispersion,} {\protect\JournalTitle{Opt. Express}} \textbf{25}, 1903--1915 (2017).

\bibitem{Liga2018a}
G.~Liga \emph{et~al.}, \enquote{A {{PMD-adaptive DBP}} receiver based on {{SNR}} optimization,} in \emph{Proc. {{OFC}},}  (2018).

\bibitem{Bitachon2020b}
B.~I. Bitachon \emph{et~al.}, \enquote{Deep learning based digital backpropagation demonstrating {SNR} gain at [...],} {\protect\JournalTitle{Opt. Express}} \textbf{28}, 29318--29334 (2020).

\bibitem{Butler2021}
R.~M. Bütler \emph{et~al.}, \enquote{Model-{{Based Machine Learning}} for {{Joint Digital Backpropagation}} and [...],} {\protect\JournalTitle{J. Lightw. Technol.}} \textbf{39}, 949--959 (2021).

\bibitem{jain_joint_2023}
P.~Jain \emph{et~al.}, \enquote{Joint {{PMD Tracking}} and {{Nonlinearity Compensation With Deep Neural Networks}},} {\protect\JournalTitle{J. Lightw. Technol.}} \textbf{41}, 3957--3966 (2023).

\bibitem{abu-romoh_equalization_2023}
M.~Abu-romoh \emph{et~al.}, \enquote{Equalization in dispersion-managed systems using learned digital back-propagation,} {\protect\JournalTitle{Opt. Continuum}} \textbf{2}, 2088 (2023).

\bibitem{Liu2023ofc}
K.~Liu \emph{et~al.}, \enquote{{{FPGA Implementation}} of {{Multi-Layer Machine Learning Equalizer}} with {{On-Chip Training}},} in \emph{Proc. OFC,}  (2023).

\bibitem{sasai_digital_2022}
T.~Sasai \emph{et~al.}, \enquote{Digital {Longitudinal} {Monitoring} of {Optical} {Fiber} {Communication} {Link},} {\protect\JournalTitle{J. Lightw. Technol.}} \textbf{40}, 2390--2408 (2022).

\bibitem{eto_location-resolved_2022}
M.~Eto \emph{et~al.}, \enquote{Location-resolved {{PDL Monitoring}} with {{Rx-side Digital Signal Processing}} in [...],} in \emph{Proc. OFC,}  (2022).

\bibitem{may_receiver-based_2022}
A.~May \emph{et~al.}, \enquote{Receiver-{{Based Localization}} and {{Estimation}} of {{Polarization Dependent Loss}},} in \emph{Proc. OECC,}  (2022).

\bibitem{takahashi_dsp-based_2023}
M.~Takahashi \emph{et~al.}, \enquote{{{DSP-based PDL Estimation}} and {{Localization}} in {{Multi-Span Optical Link Using [...]}},} in \emph{Proc. OECC,}  (2023).

\end{thebibliography}
\label{references}

\end{document}